\documentclass[aps,preprint,nofootinbib,floatfix]{revtex4}
\usepackage{epsfig}
\usepackage{graphicx}
\usepackage{multirow}
\begin{document} 

\title{\textbf{Width of a two-body coupled-channel resonance}}
\author{H.~Garcilazo} 
\email{humberto@esfm.ipn.mx} 
\affiliation{Escuela Superior de F\' \i sica y Matem\'aticas, \\ 
Instituto Polit\'ecnico Nacional, Edificio 9, 
07738 M\'exico D.F., Mexico} 

\author{A.~Valcarce} 
\email{valcarce@usal.es} 
\affiliation{Departamento de F\'\i sica Fundamental and IUFFyM,\\ 
Universidad de Salamanca, E-37008 Salamanca, Spain}

\date{\today} 

\begin{abstract}
We study the width of a two-body resonance in a coupled-channel
system. We demonstrate how the width does not come only determined 
by the available phase space for its decay to the detection channel, 
but it greatly depends on the relative position of the mass of the 
resonance with respect to the masses of the coupled-channels 
generating the state. Our results are consistent with the experimental 
observation of narrow hadrons lying well above their lowest decay threshold.
\end{abstract}

\maketitle 

\noindent
\section{Introduction} 
During the last years we have witnessed a flurry of new 
resonances in the heavy hadron spectra with unexpected properties, see the 
compilation of theoretical and experimental works in 
Refs.~\cite{Che17,Che16,Esp16,Ric16,Hos16} for a 
comprehensive overview. Apart from the fact that some of these states
have exotic flavor character or quantum numbers, in several cases another 
striking result is the lack of a clear relation between the phase space 
available for the decay in the detection channel and the width of the 
resonance. For example, in the case of the
lower LHCb pentaquark $P^+_c(4380)$~\cite{Aai15} with a mass of
$4380\, \pm \, 8 \, \pm \, 29$ MeV, it is seen to decay to the $J/\Psi p$ channel 
with a width $\Gamma = 205 \, \pm \, 18 \, \pm \, 86$ MeV. The phase space is of the order
of 345 MeV. 
Some of these states share the common feature 
that they have been suggested as possible coupled-channel hadronic resonances~\cite{Che17,Che16,Esp16,Ric16,Hos16}, 
and thus the coupled-channel dynamics may play a basic role to understand their main
features, in particular the width.

In this work we show how the width of a two-body resonance generated 
in a coupled-channel approach does not come only determined by the available 
phase space for its decay to the detection channel, but it substantially 
depends on the relative position of the mass of the resonance with respect
to the masses of the coupled-channels generating the state. Thus, one could
expect the existence of narrow hadrons lying well above their lowest decay 
threshold, in agreement with experimental observations.

\section{Formalism} 
We have studied the behavior of the width of a resonance for those cases where 
it is produced between two thresholds, thanks to a coupling between the two 
corresponding configurations within the resonance as suggested, for example, in 
Refs.~\cite{Hua13,Hua16}. For this purpose, 
we have modelled the system as a coupled-channel problem obeying  the 
non-relativistic Lippmann-Schwinger equation. Channel 1, the lower 
in mass, consists of two particles with masses $m_1$ and $m_2$, 
and channel 2, the upper in mass, is made of two particles 
with masses $m_3$ and $m_4$. The Lippmann-Schwinger equation 
in the case of $S-$wave interactions is  written as,
\begin{eqnarray}
t^{ij}(p,p';E) &=& V^{ij}(p,p')+\sum_{k=1,2} \int_0^\infty {p^{\prime\prime}}^2
dp^{\prime\prime} \nonumber \\ & \times & \!\!
V^{ik}(p,p^{\prime\prime})\,
\frac{1}{E-\Delta E \,\, \delta_{2,k}-{p^{\prime\prime}}^2/2\mu_k+i\epsilon}\, 
t^{kj}(p^{\prime\prime},p';E)\, , 
\,\,\,\,\,\,\,\, i,j =1,2,
\label{eq1} 
\end{eqnarray}
where $\mu_1=m_1 m_2/(m_1+m_2)$ and 
$\mu_2=m_3 m_4/(m_3+m_4)$ are the reduced masses of channels 1 and 2,
and $\Delta E=m_3+m_4-m_1-m_2$ with $m_3+m_4 > m_1+m_2$.
The interactions in momentum space are given by,
\begin{equation}
V^{ij}(p,p')=\frac{2}{\pi}\int_0^\infty r^2dr\; j_0(pr)V^{ij}(r)j_0(p'r),
\label{eq2} 
\end{equation}
where the two-body potentials, which are the inputs to our present study, 
consist of an attractive and a repulsive Yukawa term, i.e.,
\begin{equation}
V^{ij}(r)=-A\frac{e^{-\mu_Ar}}{r}+B\frac{e^{-\mu_Br}}{r}.
\label{eq3} 
\end{equation}
This type of parametrization is known to work rather well for the
study of two-, three-, and few-baryon systems~\cite{Mal69,Fil14} and,
thus, it is adopted here.
We have considered scenarios where a resonance exists at an 
energy $E=E_R$ such that the phase shift $\delta(E_R)=90^\circ$,
for energies between the thresholds of channels 1 
and 2, i.e., $0 < E_R < \Delta E$. The mass of the 
resonance would be given by $W_R=E_R + m_1 +m_2$.
The width of the resonance
is calculated using the Breit-Wigner formula as~\cite{Bre36,Cec14},
\begin{equation}
\Gamma (E) =\lim\limits_{E \to E_R}\, \frac{2(E_R-E)}{\text{cotg}[\delta(E)]} \, .
\label{eq4} 
\end{equation}

\section{Interacting model} 

We have parametrized the scenario depicted above,
that might be applied to several heavy resonances recently reported in the
heavy hadron spectra~\cite{Che17,Che16,Esp16,Ric16,Hos16}. Without loss 
of generality, we make use of the same thresholds recently 
considered in a more involved three-body calculation~\cite{Gar17}:
$m_1=m_2=1115.7$ MeV/c$^2$, $m_3=938.8$ MeV/c$^2$, 
and $m_4=1318.2$ MeV/c$^2$. None of the results we will later 
on discuss critically depend on the choice of the masses of the 
particles constituting the thresholds. As said
above, the interaction in the different channels is described by
Yukawa potentials consisting of an attractive and a repulsive
term. By varying the parameters one is able to control the existence of 
a bound state or a resonance and its relative position with respect to the 
thresholds, in the cases where the dynamics is dominated by two channels. 
We have chosen as starting point the set of parameters given 
in Table~\ref{t1}. 
\begin{table}[t]
\caption{Parameters of the interaction as given in Eq.~(\ref{eq3}).
$A$ and $B$ are in MeV fm, while $\mu_A$ and $\mu_B$ are in ${\rm fm}^{-1}$.} 
\begin{center}
\begin{tabular}{|p{0.5cm}cp{0.5cm}|p{0.5cm}cp{0.5cm}cp{0.5cm}cp{0.5cm}cp{0.5cm}|} 
\hline
&Channel &&&  $A$    && $\mu_A$ && $B$   && $\mu_B$  &\\ \hline
&$1\leftrightarrow 1$ &&&  $100$  && $2.68$  && $667$ && $5.81$ &\\ 
&$2\leftrightarrow 2$ &&&  $680$  && $4.56$  && $642$ && $6.73$ &\\ 
&$1\leftrightarrow 2$ &&&  $200$  && $1.77$  && $195$ && $3.33$ &\\ \hline
\end{tabular}
\end{center}
\label{t1} 
\end{table}
They are adjusted such that in a single-channel calculation, the upper channel (channel 2)
has a bound state just at threshold, while in a coupled-channel 
calculation, the full system has a bound state just at the lower 
threshold (channel 1).

\section{Results} 
\begin{figure*}[t]
\vspace*{-0.3cm}
\resizebox{10.cm}{14.cm}{\includegraphics{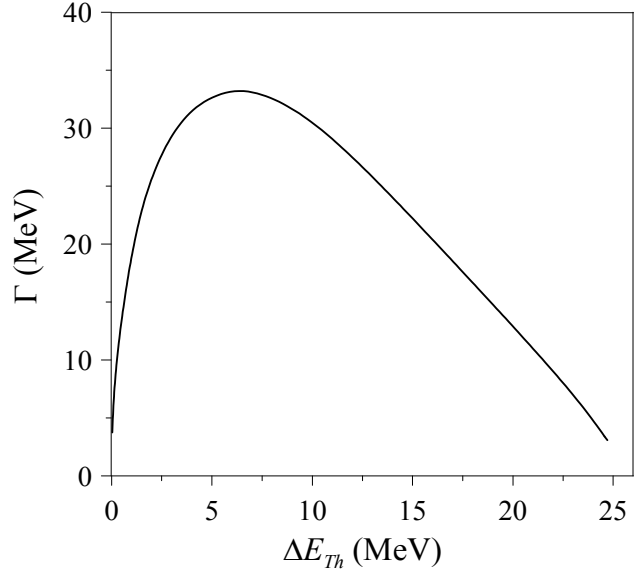}}
\vspace*{-6.0cm}
\caption{Width of the resonance, $\Gamma$, as a function of
the energy difference between its mass and 
the mass of the lower threshold generating the state, $\Delta E_{Th}=W_R-m_1-m_2$.
The upper channel is 25.6 MeV above the lower one.}
\label{fig1}
\end{figure*}

If one increases, for example, the magnitude of the repulsive term in the lower channel, 
$B(1\leftrightarrow 1)$ in Table~\ref{t1}, the bound state of the coupled-channel 
system moves up and actually becomes a resonance into the continuum.
Thus, one can study the behavior of the width of the resonance when its mass 
evolves from the lower threshold, channel 1, to the upper one, channel 2. 
The result is shown in Fig.~\ref{fig1}. 

\begin{figure*}[t]
\vspace*{-0.3cm}
\resizebox{10.cm}{14.cm}{\includegraphics{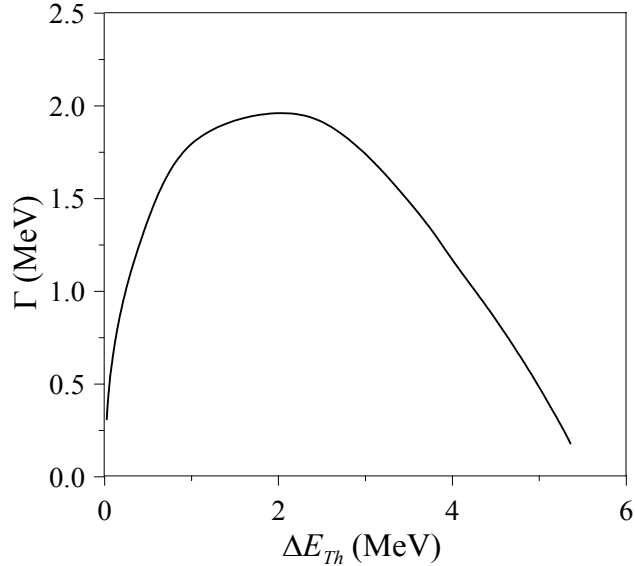}}
\vspace*{-6.0cm}
\caption{Width of the resonance, $\Gamma$, as a function of
the energy difference between its mass and the mass of the lower threshold
generating the state, $\Delta E_{Th}=W_R-m_1-m_2$.
In this case the upper channel is 5.6 MeV above the lower one.}
\label{fig2}
\end{figure*}

As one can see, the width of the resonance starts increasing quickly when
getting away from the lower threshold, but about a third of the way towards 
the upper channel, the width starts to decrease
although the phase space with respect to the threshold where the resonance is
observed still increases\footnote{Although the Breit-Wigner 
formula is not very accurate close to threshold; however, we have explicitly 
checked by analytic continuation of the S-matrix on the second Riemann sheet 
that at low energy the width follows the expected $\Gamma \sim E^{1/2}$ 
behavior, the one shown by Figures~\ref{fig1} and~\ref{fig2}.}. 
When the resonance
approaches the upper threshold, it becomes narrow  and seemingly ignores 
the existence of the lower threshold. The wave function of the $(m_3,m_4)$ bound 
state of vanishing energy has, indeed, little overlap with the $(m_1,m_2)$ configuration.
The same trend is obtained for 
different strengths of the coupling interaction in Table~\ref{t1}, channel $1\leftrightarrow 2$,
or by diminishing the repulsion in the upper channel, $B(2\leftrightarrow 2)$. 
Hence, in this region, the dynamics is dominated by the attraction 
in the upper channel and the second channel is mainly a tool for the detection.  
This mechanism is somewhat related to the 'synchronization of resonances' proposed 
by D.~Bugg~\cite{Bug08}.

The mechanism we have discussed could help to understand the narrow width of 
some experimental resonances found in the heavy hadron spectra with a large phase
space in the detection channel, whose assumed internal structure 
allow them to split into different subsystems~\cite{Che17,Esp16,Hua13,Hua16}. 
In this case the transition potential between the upper and lower channels
would come from the quark rearrangements allowed by the color structure of 
a multiquark state~\cite{Ric16,Car12}. In particular, it has been 
explained in Ref.~\cite{Car12} how hadrons with a $Q\bar Q n\bar n$ internal 
structure, where $n$ stands for a light quark and $Q$ for a heavy one, could 
split either into $(Q \bar n) - (n \bar Q)$ or
$(Q \bar Q) - (n \bar n)$. For $Q=c$ and $Q=b$, the $(Q \bar Q) - (n \bar n)$ 
threshold is lower than the $(Q \bar n) - (n \bar Q)$, the mass difference
augmenting when increasing the mass of the heavy quark (see Fig. 1 of Ref.~\cite{Car12}).
Such experimental behavior can be simply understood within quark models
with a Cornell-like potential~\cite{Isg99,Clo03}. 
Each configuration can indeed evolve into the other one,
so they cannot be considered one at a time: we are dealing with a compact
object, whose quark color quantum numbers are not separately conserved
during time evolution. This state is not a simple bound state of mesons.
Thus, the possibility of finding meson-antimeson molecules, $(Q \bar n) - (n \bar Q)$, contributing to 
the heavy meson spectra becomes more and more difficult when increasing the mass
of the heavy flavor, due to the lowering of the mass of the $(Q \bar Q) - (n\bar n)$ 
threshold\footnote{Note that the situation is completely different for $QQ\bar n \bar n$
states, due to the absence of coupled-channel dynamics~\cite{Ric16,Ric17}.}. 
This would make the system to dissociate immediately. In such cases,
the presence of an attractive meson-antimeson upper threshold together with the 
arguments we have drawn in this work, hint to a possible explanation of a narrow 
width of some of the $XYZ$ states lying close to the $(Q \bar n) - (n \bar Q)$ 
upper threshold as a meson-antimeson molecule, emphasizing the basic role of
coupled-channel dynamics~\cite{Vij14}. The situation described above would be
similar to a Feshbach resonance, where the open channel is represented
by the $(Q \bar Q) - (n\bar n)$ state that would get trapped in a molecular
state supported by the closed channel potential $(Q \bar n) - (n \bar Q)$~\cite{Bra04,Pil14}.
 
An interesting case appears when the thresholds generating the resonance
come rather close, because in this case one would find small decay widths
for the resonance in between the thresholds, that would become even smaller
when approaching any of them. This is illustrated in Fig.~\ref{fig2}, where we have
reduced the mass difference between the upper and lower  thresholds up to 5.6 MeV.
This situation may apply directly to the width of one the most elusive exotic states, 
that however has been firmly established by different collaborations and whose 
properties seem to be hardly accommodated in the quark-antiquark scheme, this 
is the $X(3872)$~\cite{Cho03}\footnote{Note, however, that there exist
some studies~\cite{Men15} finding a good fit of the data with mainly a $c\bar c$ structure.}.
Its small width, $\Gamma < 1.2$ MeV,
would fit our reasoning seeing as a coupled-channel of ${D}^0\bar D^{*0} - D^+\bar D^{*-}$,
with a mass difference of 7$-$8 MeV and being rather close to the lowest 
threshold~\cite{Bra04,Bra08}. There also could be a contribution from other channels, like
$J/\Psi \omega$~\cite{Bra08,Car09}, that it is almost degenerate with the upper
threshold $D^+\bar D^{*-}$, and thus our conclusions would remain.
Similar arguments could be handled for 
the LHCb pentaquarks, which require a careful analysis
in the models used for their study.
\begin{figure*}[t]
\vspace*{-0.3cm}
\resizebox{10.cm}{14.cm}{\includegraphics{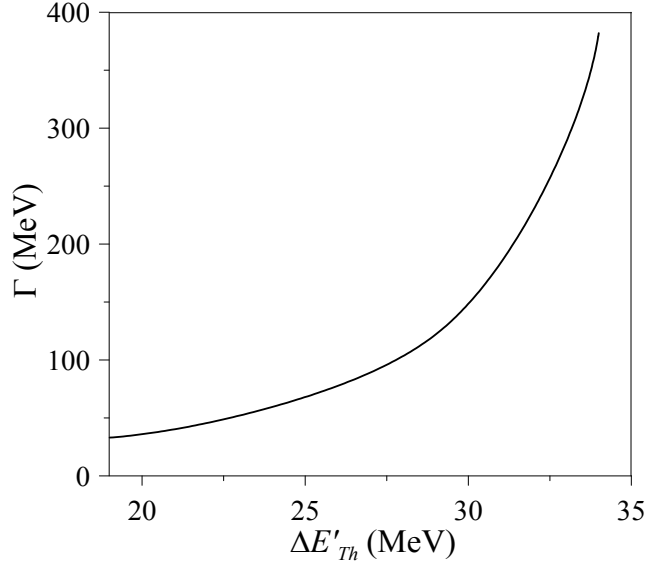}}
\vspace*{-6.0cm}
\caption{Width of the resonance, $\Gamma$, as a function of
the energy difference between its mass and the mass of the upper threshold
generating the state, $\Delta E^\prime_{Th}=m_3+m_4-W_R$, for a fixed 
energy with respect to the lower threshold, $\Delta E_{Th} = W_R-m_1-m_2=$ 6.5 MeV.}
\label{fig3}
\end{figure*}

It is obvious that the situation may become even more involved in the
case of a resonance that appears in a coupled-channel system with 
a larger number of channels. However, the described systematic of our findings
would not be modified, requiring a proper knowledge of the structure
of the resonance before estimating its decay width. The other way
around, one may conclude that an unexpected behavior of the width 
of the resonance may be indicating an important contribution of 
coupled-channel dynamics and the knowledge of the decay width in
a particular channel would hint to the upper threshold contributing to the
formation of the resonance. This has been illustrated in Fig.~\ref{fig3}, where we 
have calculated the width of the resonance for a fixed value of its mass
with respect to the lower threshold, $\Delta E_{Th} = W_R-m_1-m_2=$ 6.5 MeV, but increasing
the distance with respect to the upper threshold, $\Delta E^\prime_{Th} = m_3+m_4-W_R$. For this purpose,
we have diminished the mass of the lower channel in steps of 5 MeV, thus increasing
the distance between thresholds, $m_3+m_4-m_1-m_2$, and we have increased
$A(1\leftrightarrow 1)$ in Table~\ref{t1} in such a way that $\Delta E_{Th} =W_R-m_1-m_2=$ 6.5 MeV 
remains constant. The result is striking, being the phase space fixed for the detection channel,
the width increases when the upper threshold moves away. Thus the width provides
also with basic information about the coupled channels that
may contribute to the formation of a resonance.

\section{Conclusion} 

In this work we have studied the width of a two-body resonance in a coupled-channel
system. We have demonstrated how the width 
does not come only determined by the available phase space for
its decay to the detection channel, but it greatly depends 
on the relative position of the mass of the resonance 
with respect to the masses of the coupled-channels generating the state.
Our results are consistent with the experimental 
observation of narrow hadrons lying well above their lowest decay threshold.
Thus, they may be relevant to understand the basic features of some of 
the recently reported resonances
in the heavy hadron spectra which are suggested to be
generated by coupled-channel effects. We have also
demonstrated how an unexpected behavior of the width 
of the resonance may be indicating an important contribution of 
coupled-channel dynamics. The other way around, the observation of a 
small width in a detection channel hints to a dominant contribution 
of some upper channel to the formation of the resonance.
Hence, in this region, the dynamics is dominated by the attraction 
in the upper channel and the second channel is mainly a tool for the detection.  

Let us finally note that although the exact shape of the dependence of the
width on its position with respect to the detection channel would depend 
on the specific dynamics of the coupled-channel system,
the gross features obtained in this study might be a relevant and basic 
hint to explore the nature of some of the exotic states.

\vspace*{1cm}
\noindent
\acknowledgments 
This work has been partially funded by COFAA-IPN (M\'exico), 
by Ministerio de Econom\'\i a, Industria y Competitividad 
and EU FEDER under Contracts No. FPA2016-77177-C2-2-P and FPA2015-69714-REDT,
and by Junta de Castilla y Le\'on under Contract No. SA041U16.

\end{document}